# Nuclear Isomer Excitation in $^{229}$Th Atom by Super-Intense Laser Field

A.V. Andreev[1], A.B. Savel'ev[1], S.Yu. Stremoukhov[1,2], O.A. Shoutova[1]

[1]Faculty of Physics, Lomonosov Moscow State University, Leninskie Gory, 1, build.2, 119991, Moscow, Russia

[2]National Research Centre "Kurchatov Institute", pl. Akademika Kurchatova, 1, Moscow, 123182 Russia

Excitation of the isomeric nuclear state $^{229m}$Th in the process of thorium atom irradiation by two-color femtosecond pulse of Ti: Sa laser at the fundamental wavelength and second harmonic ($\omega+2\omega$) is analyzed. It is shown that the rate of isomeric state excitation can be enhanced significantly with respect to other nucleus excitation processes in laser plasma or by an external coherent source at the resonance wavelength. This enhancement is due to the process of "nonlinear laser nuclear excitation" based on the following. First, the atomic current associated with the motion of valence electrons in strong $\omega+2\omega$ field is a nonlinear function of the field amplitude. Secondly, the energy of photons at frequency of the Ti:Sa laser fifth harmonic coincides with the energy of the $^{229m}$Th state. Yield of this harmonic is much higher if the two-color (1$^{st}$ and 2$^{nd}$ harmonics of the fundamental) is used. Thirdly, the intensity of the field produced by valence electrons at atomic nucleus reaches the near-atomic value, which exceeds significantly the field strength associated with any other mechanism of laser plasma emission.

## I. INTRODUCTION

The thorium isomer $^{229m}$Th having an anomalously low excitation energy lying in the UV range has been attracting the attention of researchers for quite a long time [1]. Standard methods of nuclear physics give only indirect information about the properties of this unusual nuclear isomer [2-5], so that recently the estimate of the energy of this level has been changed from ~3.5 to ~8 eV [6] and at present the most reliable estimate of this energy is 7.8±0.5 eV [7]. The lifetime of an isomeric state depends essentially on the chemical environment of thorium atom, since in a non-ionized atom process of internal conversion occurs to be the most probable [8]. Recent experimental measurements gave an estimate of 7 μs for the decay through an internal conversion channel and more than 60 s through a photon channel [9, 10]. The unusual properties of this isomer give rise to a number of possible applications from a nuclear clock [10] to an optical gamma laser [11] and a quantum qubit [12]. Direct excitation of isomers from the ground state has not yet been realized, which is due to the lack of both reliable data on the excitation energy and spectrally bright tunable UV sources (with photon energy 6-15 eV).

Nuclear transitions of such small energy could be accompanied by processes in atomic shell of an atom: energy of nucleus transition can be accepted by an atomic electron (nucleus decay) or energy released in electron transition can be accepted by a nucleus (nucleus excitation). Such processes as NEEC (nuclear excitation by electron capture) [13-15], NEET (nuclear excitation by electron transition) [16,17], and BIC (bound internal conversion) [18,19] were proposed and considered for nuclear excitation. Each process mentioned above can be inversed thus giving rise to new non-photonic decay channels: internal electron conversion (IEC), inverse NEET [20] and EB (electronic bridge) [21] processes. In particular, the NEEC process is one of the candidates for realization of effective excitation of this nuclear state [13, 22-24]. Since nuclear excitation energy of the $^{229}$Th is close to energies of atomic transitions and first ionization potential of a thorium atom (6.3 eV), ionization state has crucial impact on the excitation and decay probabilities through different processes. Efficiencies of all these processes increase significantly when there is a resonance between nuclear and atomic transitions, i.e. between their energies, multipolarities, etc.

In this paper we discuss the alternative way of nucleus excitation via specific process in an electron shell under action of a strong $\omega+2\omega$ field. The idea can be easily understood from the well-known three step model of high harmonic production [25]. Here high harmonic radiation is explained from radiative recombination of an electron gaining energy after its tunnel ionization in a strong laser field. Electron wave packet accelerated back to the parent atom by the laser field after tunnel ionization can excite nuclei of this atom through NEEC, photoexcitation by high harmonic radiation, inelastic scattering at the nuclei or higher order processes in the atomic shell provided its final energy is high enough. Resonant enhancement of such an excitation can be expected if some harmonic of the driving field coincides with the nuclear excitation energy. A $\omega+2\omega$ field with crossed polarization planes provides much higher 5$^{th}$ harmonic yield and complex 3D structure of this harmonic field that can be tuned to the multipolarity of the nuclear transition.

This is valid for the excitation of the $^{229m}$Th isomeric state from the ground state of $^{229}$Th in process of illumination of thorium atom by femtosecond pulse from Ti:Sa laser. The fifth harmonic of the Ti:Sa laser is in resonance with nuclear transition from the ground to isomeric state in $^{229}$Th. Hence, resonant enhancement of the isomeric state excitation is anticipated. The target can be thorium atomic or ionized gas in a trap or metallic film of $^{229}$Th, as well as some compounds including atomic or singly ionized $^{229}$Th. Note, that quantum's energy of the fifth harmonic generated by the Ti:Sa laser radiation does not depend on the target material and can be easily tuned within 7-9 eV due to peculiar properties of the Ti:Sa medium.

## II. BASIC DESCRIPTION

To tackle the problem, we applied recently developed approach [26, 27] using orthonormal basis of eigenfunctions of boundary value problem for "an atom in the external field". It has been proposed and verified for analysis of the problems of light-atom interactions in the strong laser fields. The eigenfunctions of this complete orthonormal basis are time-dependent. As a result, the wave function of any atomic electron evolves in time even if it is in eigenstate. Harmonics of the driving laser field appear in the response field when this temporal dependency is periodic [28]. The energy for nucleus excitation and harmonics generation is equal to the work produced by the laser field on the electron movement in the intra-atomic field.

The matrix elements of nuclear transitions depend on the mutual orientation of nucleus angular momentum and the polarization of the driving e.m. field. At the same time, the polarization of a field produced by an electron at nucleus depends on the mutual orientation of the polarization vector of driving laser field and the angular momentum of an atomic electron. The rates of radiative excitation of $^{229}$Th in transition 3/2+ - 5/2+ associated with the magnetic dipole (M1) and/or electric quadrupole (E2) transitions have been repeatedly calculated (see [10]). Therefore, there is no need to discuss these calculations in detail. As a result, we shall not discuss this problem but elaborate on the calculation of the vector state of a resonant e.m. field at nucleus under given polarization of transition and laser field and orientation of atomic electron angular momentum.

We assume that at the initial moment both the nuclear and electronic subsystems of the atom are in the ground state. The developed mathematical model takes into account the following features. First, the 6d$_2$ state is the ground state of the valence electron band of the thorium atom or ion, so when the atomic thorium or the compounds including singly ionized thorium are excited, it is necessary to take into account the nonzero value of the angular momentum of the valence electrons. Since, as shown in [26, 27], the rate of radiative excitation depends on the mutual orientation of the polarization vector of the exciting laser field and the angular momentum of the atom. Secondly, the routine calculations of the atomic response spectrum, including the field of high harmonics, deal with the far field components. Generally, in the solid-density target, the excitation of nuclei can be due to the far field of harmonics generated by neighboring atoms. But the nucleus is evidently localized in the near-field zone of the harmonics generated by the electrons of the native atom. Therefore, the calculation of the atom response field should be performed without the use of the far-field approximation. Thirdly, the used theory has vectorial nature, so it allows calculating not only the amplitude of the field acting on the core, but also its polarization.

## III. THEORY

The spectrum of harmonics generated by intense laser pulse is usually calculated for the far field zone [29]

$$\vec{E}(\vec{r},\omega) = -\frac{i\omega}{c^2}\frac{1}{r}\exp\left(i\frac{\omega}{c}r\right)\left[\vec{n}\left[\vec{n}\vec{J}(\omega)\right]\right], \quad (1)$$

where r is the distance from the radiating atom to the observation point, $\vec{J}(\omega)$ is the spectral density of the atomic current. This formula describes the spectrum of the transverse response field, the amplitude of which decreases along with the distance from the radiating atom as 1/r. It is not difficult to estimate the magnitude of this field. In the problems of interaction of an atom with strong laser fields, the intensity of the intra-atomic field is the characteristic unit of the field intensity, i.e. the field acting on an electron in a hydrogen atom

$$E_a = \frac{e}{a_B^2}, \quad (2)$$

where $a_B$ is Bohr radius.

Let us compare the atomic response field magnitude (1) with the intra-atomic one. The value of field (1) can be approximated as follows

$$E(r,\omega) \approx e\frac{\omega}{c}\frac{1}{r}\frac{\langle v\rangle}{c} \approx \frac{e}{\lambda r}\frac{\langle v\rangle}{c}, \quad (3)$$

where $\langle v\rangle$ is the mean velocity of atomic electron motion in the external laser field. As far as we are interested in the probability of excitation of the atomic nucleus by radiation generated by moving atomic electrons we can put $r \approx a_B$. In the case of strong laser fields, the ratio $\langle v\rangle/c$ does not play any decisive role because the values $\langle v\rangle/c = 0.1-0.9$ can be easily achieved. Comparing (2) and (3), we can easily see that the field strength at nucleus depends significantly on the nuclear transition energy. Indeed, if we are discussing the possibility of exciting the nuclear levels with an energy which coincides with the photon energy in the visible range or close to it (when $\lambda \gg a_B$), the field strength (1) is much smaller than the intra-atomic field strength in spite of the fact that $r \approx a_B$.

It is clear that the model based on the calculation of the field in the far zone (1) is inapplicable in calculations of nucleus excitation probability in the case when the exciting photon is born as a result of the atomic electron motion in the external laser field. Therefore, we calculate the amplitude of the response field of an atomic electron at an arbitrary point in space and for an arbitrary value of the velocity of its motion in an external laser field.

The solutions of the wave equations for the vector and scalar potentials have the form of

$$\vec{A}(\vec{r},t) = \frac{1}{c}\int\frac{\vec{j}(\vec{r}',t-|\vec{r}-\vec{r}'|/c)}{|\vec{r}-\vec{r}'|}dV', \quad (4a)$$

$$\varphi(\vec{r},t) = \int\frac{\rho(\vec{r}',t-|\vec{r}-\vec{r}'|/c)}{|\vec{r}-\vec{r}'|}dV'. \quad (4b)$$

The vector of the electric field strength is given by

$$\vec{E}(\vec{r},t) = -\frac{1}{c^2}\int\frac{1}{|\vec{r}-\vec{r}'|}\frac{\partial\vec{j}(\vec{r}',t-|\vec{r}-\vec{r}'|/c)}{\partial t}dV' -$$
$$-\int\frac{\partial}{\partial\vec{r}}\left(\frac{\rho(\vec{r}',t-|\vec{r}-\vec{r}'|/c)}{|\vec{r}-\vec{r}'|}\right)dV'. \quad (5)$$

Performing differentiation in the second term, we obtain

$$\vec{E}(\vec{r},t) = -\frac{1}{c^2}\int\frac{1}{|\vec{r}-\vec{r}'|}\frac{\partial\vec{j}(\vec{r}',t')}{\partial t}dV' +$$
$$+\frac{1}{c}\int\frac{\vec{r}-\vec{r}'}{|\vec{r}-\vec{r}'|^2}\frac{\partial\rho(\vec{r}',t')}{\partial t}dV' + \int\frac{\vec{r}-\vec{r}'}{|\vec{r}-\vec{r}'|^3}\rho(\vec{r}',t')dV', \quad (6)$$

where $t' = t - |\vec{r}-\vec{r}'|/c$.

As far as the potentials (4a-b) are solutions of the wave equations in the Lorentz gauge, then according to (4a-b) we get

$$\frac{1}{|\vec{r}-\vec{r}'|}\frac{\partial\rho(\vec{r}',t')}{\partial t} + \frac{\partial}{\partial\vec{r}}\left(\frac{\vec{j}(\vec{r}',t')}{|\vec{r}-\vec{r}'|}\right) = 0. \quad (7)$$

Substituting (7) in (6), we get

$$\vec{E}(\vec{r},t) = -\frac{1}{c^2}\int\frac{1}{|\vec{r}-\vec{r}'|}\left[\frac{\partial\vec{j}(\vec{r}',t')}{\partial t} - \vec{n}'\left(\frac{\partial\vec{j}(\vec{r}',t')}{\partial t}\vec{n}'\right)\right]dV' +$$
$$+\frac{1}{c}\int\frac{(\vec{n}'\vec{j}(\vec{r}',t'))\vec{n}'}{|\vec{r}-\vec{r}'|^2}dV' + \int\frac{\vec{n}'\rho(\vec{r}',t')}{|\vec{r}-\vec{r}'|^2}dV', \quad (8)$$

where $\vec{n}' = \frac{\vec{r}-\vec{r}'}{|\vec{r}-\vec{r}'|}$.

Thus, the first term in (8) determines the intensity of the transverse electromagnetic field at the observation point as follows

$$\vec{E}_\perp(\vec{r},t) = \frac{1}{c^2}\int \frac{1}{|\vec{r}-\vec{r}'|}\left[\vec{n}'\left[\vec{n}'\frac{\partial \vec{j}(\vec{r}',t-|\vec{r}-\vec{r}'|/c)}{\partial t}\right]\right]dV'. \quad (9)$$

The remaining two terms in (8) determine the strength of the longitudinal electric field at the observation point.

The division of the field into longitudinal and transverse components is due to the fact that the electric field vector can have a non-zero projection both on the direction connecting the electron orbit coordinate with the atomic nucleus and on the two perpendicular directions. In the recent paper [30, 31], we have shown that a longitudinal atomic current arises also in a single atom if the angular momentum of the atomic state is not equal to zero. The presence of a longitudinal component of atomic current changes substantially the polarization properties of the atomic response field and, as can be seen from the above formulas, changes significantly the value of the probability of excitation of the nucleus. Indeed, instead of (3) we get

$$E(r,\omega) \approx \frac{\langle v\rangle}{c}E_a. \quad (10)$$

We see that the nucleus excitation probability increases substantially and becomes comparable with the probability of excitation of the nucleus in the laser field of the near-atomic strength. Note that external production of the fifth harmonic of the Ti:Sa laser radiation amounts to pulse energies below 1 µJ [32], that can produce the field strength below $0.1E_a$ even at tight focusing conditions.

As we noted above, the transition with energy of 7.8 eV in $^{229}$Th can be magnetic dipolar, so we write out the equation for the amplitude of the magnetic field in the near field zone

$$\vec{B}(\vec{r},t) = \frac{1}{c}\int \frac{1}{|\vec{r}-\vec{r}'|}\left[\frac{\partial \vec{j}(\vec{r}',t')}{\partial t}\vec{n}'\right]dV' + \int \frac{[\vec{j}(\vec{r}',t')\vec{n}']}{|\vec{r}-\vec{r}'|^2}dV'. \quad (11)$$

Similarly to the previous discussion, it is clear that in the near-field zone the second term is dominated. Its amplitude at nucleus exceeds greatly the field amplitude determined by the first term.

## IV. NUMERICAL CALCULATIONS

The series of computer simulations on the atomic current induced in the electron shell of the thorium atom by the femtosecond pulse of Ti: Sa laser has been made to determine the properties of the fifth harmonic response field at atomic nucleus. As we have mentioned above in our simulations we use the non-perturbative vectorial theory of high optical harmonic generation [26, 27], so we have calculated not only the amplitude of the fifth harmonic field, but its polarization state at given mutual orientation of atomic angular momentum, driving laser field polarization and laser pulse propagation direction (see for details [30,31]). In numerical simulations the model structure of the $^{229}$Th atom included the following discrete spectrum states: 6d, 7s, 5f, 7p. Calculations were carried out for the case of the ω+2ω orthogonally polarized laser waves at the fundamental frequency and the second harmonic of the Ti:Sa laser – the simplest scheme of ω+2ω field [33]; the intensity of both components of the ω+2ω field are assumed to be ~ $10^{12}$ W/cm2. We have chosen the ω+2ω field for the simulation for reasons: firstly, to increase efficiency of the fifth harmonic generation [33], secondly, to increase amplitude of the longitudinal field generation [30]. More importantly, having all three components of the radiation with more or less the same field strength makes it possible to easily adjust to the 3/2+ - 5/2+ transition and to more effectively excite the nucleus. In numerical experiments, the three components of the atomic current on the Cartesian set of coordinates are calculated. It is assumed that the x and y-axes coincide with the direction of the polarization vectors of the driving field at the frequency of Ti: Sa laser ($j_1$) and its second harmonic ($j_2$), respectively, whereas the z-axis coincides with the direction of the ω+2ω wave propagation ($j_3$) [29].

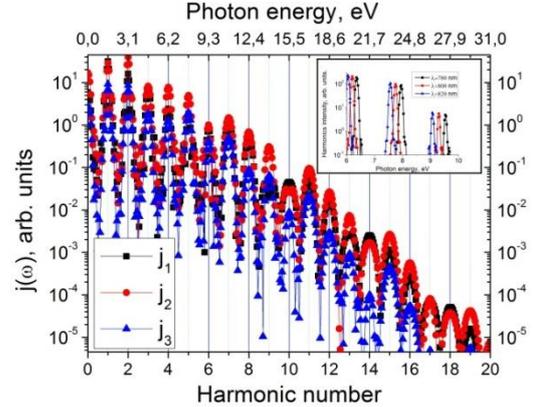

Fig. 1. Spectra of atomic current projections for the case of $^{229}$Th atom interaction with the ω+2ω field formed by the fundamental and the second harmonics of the Ti:Sa laser (λ=800 nm) with the intensities of the components of ~ $10^{12}$ W/cm²; the Euler angles $\theta_0=\psi_0=$ 0.5, $\varphi_0=0$. *Inse*t: the variation in the position of some selected harmonics in spectra under variation of the driving laser field frequency.

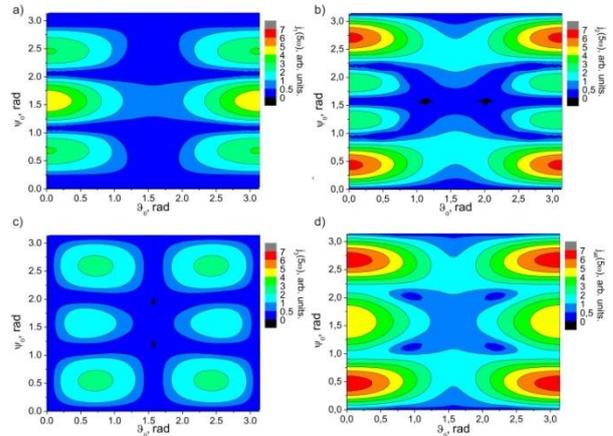

Fig. 2. Projections of atomic current $j_1$ (a), $j_2$ (b), $j_3$ (c) and overall atomic current (d) as functions of $\theta_0$, $\psi_0$ ($\varphi_0=0$), calculated for $^{229}$Th atom interacting with the ω+2ω field at fundamental frequency and the second harmonic of the Ti:Sa laser (λ=800 nm) with the intensities of ~ $10^{12}$ W/cm².

The polarization of the fifth harmonic field at nucleus strongly depends on the propagation direction of the driving field and the polarizations of ω+2ω components. So, if the thorium target is polarized, then by varying the propagation direction of the driving field we can control the amplitude and the polarization of the fifth harmonic field at nucleus. The spectra of the atomic current projections calculated for some specific values of the Euler angles ($\theta_0$, $\psi_0$, $\varphi_0$), connecting the axes of atomic configurational space and coordinate set associated with configuration of the ω+2ω field [30]) are shown in Fig. 1. The spectrum of the three atomic current components contains even and odd field harmonics, including the 5th harmonic. The inset in Fig.1 shows that by varying the carrier frequency of the Ti:Sa laser we can easily vary the wavelength of the fifth harmonic, i.e. we can easily tune the energy of the fifth harmonic quantum to the resonance with the energy of the $^{229m}$Th state.

The map of the atomic current amplitude as a function of angles $\theta_0$ and $\psi_0$ ($\varphi_0=0$ for simplicity) is shown in Fig.2. In Fig. 2 (a-c), the Cartesian components of atomic current are shown; the amplitude of total current is shown in Fig. 2 d. Notice that according to the equation (8), the amplitude of the fifth harmonic is proportional to the atomic current amplitude. From the analysis of Fig. 2 one can see that the field strength of the fifth harmonic is a non-linear function of angles $\theta_0$ and $\psi_0$. Hence, we can easily control the rate of nucleus excitation by varying the parameters of the driving $\omega+2\omega$ field (direction of propagation and polarization the field components).

## V. CONCLUSIONS

We have studied possibility of the isomeric state $^{229m}$Th excitation under action at a target containing neutral or singly ionized thorium atoms by intense two- color laser field, which consists of copropagating orthogonally polarized waves at fundamental frequency of the Ti: Sa laser and its second harmonic. The results of simulations have shown the possibility of high efficiency of excitation. The high excitation efficiency is due to the following reasons. First, the excitation energy of the nucleus coincides with the energy of photons at the frequency of the fifth harmonic of the Ti:Sa laser. Secondly, the fifth harmonic field strength at nucleus depends non-linearly on the mutual orientation of the laser field and atomic angular momentum. Thirdly, the amplitude of the field produced by valence electrons at atomic nucleus reaches the near-atomic values which exceed significantly the field strength associated with any other mechanisms of laser plasma emission or external coherent optical source. Indeed, let us compare the field strength produced by the oscillating atomic electron on its own atom and the neighboring atoms. Assuming that the mean distance between the atoms or ions in plasma is about $d \approx 10$ Å for the ratio of transversal field the field strength produced by oscillating atomic electron on its own nucleus (10) to the one produced on neighboring atoms (3) we get $E_{own}/E_{neighb} \approx 10^4 - 10^5$.

We performed calculations in the framework of the vector theory of harmonics generation which opens up new possibilities for spectroscopic studies. Indeed, the direction of the field strength vector on nucleus at the frequency of the fifth harmonic depends on the mutual orientation of the laser field vector and the angular momentum of atomic electrons. Therefore, if an atom or ion being introduced into some matrix or structure is in a polarized state, then the direction of the field acting on nucleus can be easily controlled by a variation of the propagation direction of the laser wave and its polarization (see Fig.2).

Even more interesting, the spectroscopic information can be obtained by placing the sample under study in a magnetic field (to polarize the state of the nuclei). It should be noted that when the duration of laser pulses is in femtosecond range, then only the state of the nucleus at the initial moment is important for further dynamics of the system. Indeed, for a given position of the detector in the observation chamber, the excitation efficiency depends on the mutual orientation of the laser beam, its polarization, the position of target, and the direction of the magnetic field. Therefore, the detector's readings will depend on these parameters, even if the target is absolutely destroyed by the laser field action and the atoms are deposited on the walls of the chamber.

The proposed mechanism of nucleus excitation via a nonlinear response of the electronic shell can be called the the "nonlinear laser nuclear excitation" or NLEX.

## ACKNOWLEDGMENTS

The work was partially supported by the RFBR under Projects Nos. 18-02-00528, 18-02-00743, 16-52-10012.